\documentclass[onecolumn,aps,pra,amsfonts,amssymb]{revtex4}
\usepackage{graphicx}
\usepackage{amsmath}
\usepackage{bm}

\usepackage{amssymb}

\begin{document}

\title{On superluminal propagation and information velocity}
\date{\today}
\author{Akhila Raman} \affiliation{University of California at Berkeley, CA-94720. Email: akhila.raman@berkeley.edu. }


\begin{abstract}
This paper examines some of the recent experiments on superluminal propagation. It is well known that Sommerfeld and Brillouin analyzed a rectangular sinusoidal  signal propagating through a dispersive medium and derived  expressions to describe the precursors and the main signal.  In this paper, the impulse response of this dispersive medium is derived as exact expression using Taylor series expansion and output signal for any causal input signal is shown to be zero for time $t$ less than vaccum transit time  and implications for superluminal information velocity is analyzed.\\

\end{abstract}

\maketitle

\section{Introduction}

Let us start with the well known formulation used by Sommerfeld and Brillouin$^{3}$ for a rectangular sinusoidal  signal propagating through a dispersive medium and using the modern notations$^{1}$ and ignoring reflections at the interface, we can write as follows.

\begin{eqnarray*}
u(x,t)=\frac{1}{2\pi} \int_{-\infty}^{\infty} A(\omega) e^{i k(\omega)x-i \omega t} d\omega \\
k(\omega) = \frac{\omega n(\omega)}{c} \\
A(\omega)=  \int_{-\infty}^{\infty} u(0,t) e^{i \omega t}  dt \\
u(0,t)= rect(\frac{t-\frac{T}{2}}{T}) \sin(\omega_{0} t) \\
A(\omega) = \frac{1}{2i} [A_{0}(\omega+ \omega_{0}) - A_{0}(\omega - \omega_{0})] \\
A_{0}(\omega) = \frac{2}{\omega} \sin(\omega \frac{T}{2}) e^{i \omega \frac{T}{2}}
 \end{eqnarray*}
\begin{equation}  \end{equation}

where $n(\omega)$ is the refractive index of this dispersive medium, $k(\omega)$ is the wave number and $c$ is the speed of light in vaccum. Let $L$ be the length of this dispersive medium. At $x=0$, we have the input signal is given by $x(t)=u(0,t)=\frac{1}{2\pi} \int_{-\infty}^{\infty} A(\omega) e^{-i \omega t} d\omega = rect(\frac{t-\frac{T}{2}}{T}) \sin(\omega_{0} t)$. At $x=L$, we get the output signal given by  $y(t)=u(L,t)=\frac{1}{2\pi} \int_{-\infty}^{\infty} A(\omega) e^{i k(\omega)L-i \omega t} d\omega = \frac{1}{2\pi} \int_{-\infty}^{\infty} A(\omega) e^{i \frac{\omega n(\omega)L}{c}} e^{-i \omega t} d\omega $. Considering the dispersive medium as a Linear Time Invariant system(LTI), we can write the output signal as the convolution of the input signal and impulse response given by $y(t)=\int_{-\infty}^{\infty} x(\tau) h(t-\tau) d\tau$ where $h(t)$ is the impulse response of the medium. Applying Fourier Transform to this equation we get $Y(\omega)=X(\omega)H(\omega)$ where $H(\omega)$ is the frequency response of the medium given by $H(\omega)= \int_{-\infty}^{\infty} h(t) e^{i \omega t} dt = \frac{Y(\omega)}{X(\omega)} = e^{i \frac{\omega n(\omega)L}{c}}$. Let us denote $t_{0}= \frac{L}{c}$ as the vaccum transit time. We have $H(\omega)=  e^{i \omega n(\omega)t_{0}}$.

It is well known that the refractive index
of the dispersive absorbing medium is given by $n(\omega)=[1+ \frac{\omega_{p}^{2}}{\omega_{0}^{2}-\omega^{2}-i \gamma \omega}]^{\frac{1}{2}}$   and $\omega_{p}, \omega_{0}, \gamma$ refer to the plasma frequency, resonance frequency and damping constant respectively.  Given the recent experiments$^{4,5}$ involving subluminal and superluminal light velocity in absorbing media and gain media, it may be of interest to derive an exact expression for the signal propagating through such media.  In this paper, the impulse response of this absorbing medium  is  derived as exact expressions using Taylor series expansion for $|\frac{\omega_{p}^{2}}{\omega_{0}^{2}-\omega^{2}-i \gamma \omega} |< 1$. For Gain media, $\omega_{p}^{2}$ is merely replaced by $-\omega_{p}^{2}$. \\

\section{Section 2 \label{Sec2}}

Let the refractive index be given by $n(\omega)=[1+\chi(\omega)]^{\frac{1}{2}}$ where $\chi(\omega)=\frac{\omega_{p}^{2}}{\omega_{0}^{2}-\omega^{2}-i \gamma \omega}$ denotes electric susceptibility. We can expand $n(\omega)$ as follows for the case $|\chi(\omega)| < 1$ in the range $-\infty \leq \omega \leq \infty $.  $|\chi(\omega)| < 1$  is possible when $ \omega_{p}^{2} < \gamma \omega_{0}$. 

\begin{equation} n(\omega)= \sum_{r=0}^{\infty} \binom{\frac{1}{2}}{r}  [ \frac{\omega_{p}^{2}}{\omega_{0}^{2}-\omega^{2}-i \gamma \omega} ]^{r} \end{equation}

Hence we can expand the medium frequency response $H(\omega)$ in Taylor's series as follows:

\begin{eqnarray*}H(\omega)= e^{i\omega n(\omega) t_{0}} = \sum_{k=0}^{\infty} \frac{1}{!k} [i\omega n(\omega) t_{0}]^{k} 
= 1 + i\omega n(\omega) t_{0} + \frac{1}{!2}(i\omega n(\omega) t_{0})^{2} +.... \end{eqnarray*}
\begin{equation}  \end{equation}

Rearranging the even terms and odd terms of this Taylor series expansion and
writing $\frac{\omega_{p}^{2}}{\omega_{0}^{2}-\omega^{2}-i \gamma \omega} = \frac{A}{(\omega - \omega_{1})(\omega - \omega_{2})}$ where $A=-\omega_{p}^{2}$, $\omega_{1}=-i \frac{\gamma}{2} + \sqrt{\omega_{0}^{2}- \frac{\gamma^{2}}{4}}$, $\omega_{2}=-i \frac{\gamma}{2} - \sqrt{\omega_{0}^{2}- \frac{\gamma^{2}}{4}}$, $y=n^{2}(\omega)=[1+ \frac{A}{(\omega - \omega_{1})(\omega - \omega_{2})}]$, we get

\begin{eqnarray*} H(\omega)= [1 + \frac{1}{!2}(i\omega t_{0})^{2}  y + \frac{1}{!4}(i\omega t_{0})^{4}  y^{2} + ...] 
+ y^{\frac{1}{2}} [ (i\omega t_{0}) + \frac{1}{!3}(i\omega t_{0})^{3}  y + \frac{1}{!5}(i\omega t_{0})^{5}  y^{2}  + ...    ]  \end{eqnarray*}
\begin{equation}  \end{equation}

Using the fact that $y^{\frac{1}{2}} = n(\omega)= \sum_{r=0}^{\infty} \binom{\frac{1}{2}}{r}  [ \frac{A^{r}}{(\omega - \omega_{1})^{r}(\omega - \omega_{2})^{r}} ]$ and the fact that 
$y^{k}= \sum_{n=0}^{k} \binom{k}{n}  \frac{A^{n}}{(\omega - \omega_{1})^{n}(\omega - \omega_{2})^{n}} $, we get

\begin{eqnarray*}
   H(\omega)= \sum_{k=0}^{\infty} \frac{(-\omega^{2} t_{0}^{2})^{k}}{!(2k)} [\sum_{n=0}^{k} \binom{k}{n}  \frac{A^{n}}{(\omega - \omega_{1})^{n}(\omega - \omega_{2})^{n}}]  
  + (i \omega t_{0}) \sum_{k=0}^{\infty} \frac{(-\omega^{2} t_{0}^{2})^{k}}{!(2k+1)} \sum_{r=0}^{\infty} \binom{\frac{1}{2}}{r} [\sum_{n=0}^{k} \binom{k}{n}  \frac{A^{n+r}}{(\omega - \omega_{1})^{n+r}(\omega - \omega_{2})^{n+r}}]
\end{eqnarray*}

\begin{equation} \end{equation}

For $t < t_{0}$, it is easy to show that $ h(t)=\frac{1}{2\pi} \int_{-\infty}^{\infty} H(\omega) e^{-i \omega t} d\omega = 0$. Using $n(\omega)=1 +  \sum_{r=1}^{\infty} \binom{\frac{1}{2}}{r}  [ \frac{A^{r}}{(\omega - \omega_{1})^{r}(\omega - \omega_{2})^{r}} ] = 1 + n_{1}(\omega)$, we have $h(t)=\frac{1}{2\pi} \int_{-\infty}^{\infty} H(\omega) e^{-i \omega t} d\omega = \frac{1}{2\pi} \int_{-\infty}^{\infty} e^{i\omega n_{1}(\omega) t_{0}} e^{-i \omega (t- t_{0})} d\omega  $. Given that $e^{i\omega n_{1}(\omega) t_{0}} e^{-i \omega (t- t_{0})}$ is analytic in the upper-half complex-plane with no singularities for $t<t_{0}$, it is easy to show that $h(t)=0$ for $t<t_{0}$. \\

We wish to find $ h(t)=\frac{1}{2\pi} \int_{-\infty}^{\infty} H(\omega) e^{-i \omega t} d\omega$ by using contour integration and Cauchy's theory of residues for $t>=t_{0}$.  Given that $H(\omega)e^{-i \omega t}$ is analytic in the lower-half complex-plane except at singularities at $\omega=\omega_{1}$ and $\omega=\omega_{2}$(from Eq.5), using Cauchy's Residue theorem, we can compute the residues at these singularities as follows. Let us examine factors of the form $\frac{\omega^{a}}{(\omega - b)^{n}}$ which figure in the above equation. We can write its derivatives $0,1,2,..n$ with respect to $\omega$ as follows.

\begin{eqnarray*}f_{0}(\omega, b, n, a) = \frac{\omega^{a}}{(\omega - b)^{n}} \\
f_{1}(\omega, b, n, a) = \frac{-n \omega^{a}}{(\omega - b)^{n+1}} +  \frac{a \omega^{a-1}}{(\omega - b)^{n}} \\
f_{2}(\omega, b, n, a) = \frac{n(n+1) \omega^{a}}{(\omega - b)^{n+2}} + \frac{-2na \omega^{a-1}}{(\omega - b)^{n+1}} +  \frac{a(a-1) \omega^{a-2}}{(\omega - b)^{n}} \\
f_{3}(\omega, b, n, a) = \frac{-n(n+1)(n+2) \omega^{a}}{(\omega - b)^{n+3}} + \frac{3n(n+1) a \omega^{a-1}}{(\omega - b)^{n+2}}  + \frac{-3na(a-1) \omega^{a-2}}{(\omega - b)^{n+1}} +  \frac{a(a-1)(a-2) \omega^{a-3}}{(\omega - b)^{n}}  \end{eqnarray*}

\begin{eqnarray*} f_{n-1}(\omega, b, n, a) =\sum_{r=0}^{n-1} 
(\frac{(-1)^{n-r-1} \binom{n-1}{r} \omega^{a-r} \prod_{l=0}^{n-r-2}(n+l)\prod_{l=0}^{r-1}(a-l)}{(\omega - b)^{2n-r-1}} ) \end{eqnarray*}
\begin{equation} \end{equation}

Using this, we can write the impulse response of the absorbing medium for $t >= t_{0}$ as follows, using the theory of residues. Using $ h(t)=\frac{1}{2\pi} \int_{-\infty}^{\infty} H(\omega) e^{-i \omega t} d\omega$ and expanding $e^{-i \omega t}$ in Taylor series, we have

\begin{eqnarray*}
   h(t) = \frac{1}{2 \pi} \sum_{m=0}^{\infty} \frac{(-it)^{m}}{!m} [ \sum_{k=0}^{\infty} \frac{(-t_{0}^{2})^{k}}{!(2k)} [\sum_{n=0}^{k} \binom{k}{n}  G(2k+m,n)] + (i  t_{0}) \sum_{k=0}^{\infty} \frac{(-t_{0}^{2})^{k}}{!(2k+1)} \sum_{r=0}^{\infty} \binom{\frac{1}{2}}{r} [\sum_{n=0}^{k} \binom{k}{n}  G(2k+m+1,n+r)] ] \\ t>=t_{0} 
\end{eqnarray*}

 where $ G(2k+m,n), G(2k+m+1,n+r)$ are the residues.

\begin{eqnarray*}
  G(a,n')= 2 \pi i \frac{A^{n'}}{!(n'-1)} [ f_{n'-1}(\omega_{1}, \omega_{2}, n', a) + f_{n'-1}(\omega_{2}, \omega_{1}, n', a) ] 
  \end{eqnarray*}

\begin{eqnarray*}
f_{n'-1}(\omega, b, n', a) = \sum_{r=0}^{n'-1} 
(\frac{(-1)^{n'-r-1} \binom{n'-1}{r} \omega^{a-r} \prod_{l=0}^{n'-r-2}(n'+l)  \prod_{l=0}^{r-1}(a-l)}{(\omega - b)^{2n'-r-1}} ) 
\end{eqnarray*}

\begin{equation} \end{equation}

Let us use the expression derived for $h(t)$ in the above section to develop  expressions for the output signal $y(t)=u(L,t)$. We know that
$x(t)=u(0,t)= rect(\frac{t-\frac{T}{2}}{T}) \sin(\omega_{0} t)$ and $y(t)=\int_{-\infty}^{\infty} h(\tau) x(t-\tau)  d\tau$ and $h(t)=0$ for $t< t_{0}$. Hence we can see that $y(t)=0$ for $t< t_{0}$. Hence superluminal propagation and superluminal information velocity seems not possible in absorbing medium, for $|\chi(\omega)| < 1$.
For Gain media, $\omega_{p}^{2}$ is merely replaced by $-\omega_{p}^{2}$ and superluminal propagation and superluminal information velocity seems not possible.

\section{ Section 3 \label{Sec4}}

We can expand the refractive index $n(\omega)=[1+\chi(\omega)]^{\frac{1}{2}}$ where $\chi(\omega)=\frac{\omega_{p}^{2}}{\omega_{0}^{2}-\omega^{2}-i \gamma \omega}$ as follows for the case $|\chi(\omega)| >= 1$. Let $z=\chi(\omega)$ and we can express $(1+ z)^{\frac{1}{2}}$ for $|z| >= 1$ as follows$^{6}$.\\

It is well known that the Taylor series expansion of  $(1+ z)^{A}$ is given by

\begin{equation} 
(1+ z)^{A}= \sum_{n=0}^{\infty} \binom{A}{n}  z^{n} \end{equation}

 for $|z|<1$ where  $\binom{A}{n}$ is the binomial choose function. It is well known that 
the series expansion does not converge for $|z|>1$ where A is a real number which is not equal to zero or a positive integer.

We could obtain a limited series expansion for $|z|>1$ by writing the above expression as follows
\begin{eqnarray*}(1+ z)^{A}  =  (1+ \frac{z}{2} + \frac{z}{2}  )^{A} =  (1+ \frac{z}{2}  )^{A}  (1+ \frac{ \frac{z}{2}}{ (1+ \frac{z}{2}) }  )^{A} \\ = (1+ \frac{z}{2}  )^{A}  (1+ \frac{z}{ z+ 2 }  )^{A}  \end{eqnarray*}  
\begin{equation}   \end{equation}

The second term in the above equation has a convergent series representation, given that $|\frac{z}{ z+ 2 }| < 1$. If   $|\frac{z}{2}|>1$, we can write
\begin{equation}    (1+ \frac{z}{2})^{A}  =   (1+ \frac{z}{4}  )^{A}  (1+ \frac{z}{ z+ 4 }  )^{A}    \end{equation}

Repeating this procedure iteratively, if   $m_{0}$ is the minimum value for which $|\frac{z}{2^{m_{0}}}|<1$, we can write

\begin{equation}     (1+ z)^{A}  =   (1+ \frac{z}{2^{m_{0}}})^{A} \prod^{m_{0}}_{r=1}  (1+ \frac{z}{ z+   2^{r} }  )^{A}    \end{equation}

Each of the terms in the above product of terms has a convergent series representation. Given that we can write the convergent series expansion for each of the terms above as 
$(1+ \frac{z}{2^{m_{0}}})^{A}= \sum_{n=0}^{\infty} \binom{A}{n}  (\frac{z}{2^{m_{0}}})^{n}$
and $(1+ \frac{z}{ (z+   2^{r}) }  )^{A} =  \sum_{m=0}^{\infty} \binom{A}{m} (\frac{z}{ (z+   2^{r}) })^{m}$, where $\binom{A}{n}$ represents the Choose function[2],  we have the \textbf{series expansion for $(1+ z)^{A}$ expressed 
as a product of convergent series, which converges for } $|z|>1$  as follows:

\begin{eqnarray*}  (1+ z)^{A}   =  [ \sum_{n=0}^{\infty} \binom{A}{n}  (\frac{z}{2^{m_{0}}})^{n}     ] [ \prod^{m_{0}}_{r=1}  \sum_{m=0}^{\infty} \binom{A}{m}  [\frac{z}{ z+   2^{r}}]^{m} ] \end{eqnarray*}  
\begin{equation} \end{equation}

Now we can substitute $z = \chi(\omega)$ and $A=\frac{1}{2}$ in the above expression
and obtain the series expansion of $n(\omega)$ and substitute it for $y^{\frac{1}{2}}$ in Eq.4 and
use the procedure outlined in Section 2  to derive similar expression for the impulse response $h(t)$ and $y(t)$.\\

\section{Conclusions\label{Sec7}}
Sommerfeld and Brillouin analyzed a rectangular sinusoidal  signal propagating through a dispersive medium and derived  expressions to describe the precursors and the main signal.  In this paper, the impulse response of this dispersive medium is derived as exact expression using Taylor series expansion and output signal for any causal input signal is shown to be zero for time $t$ less than vaccum transit time  and hence superluminal propagation and superluminal information velocity seems not possible in absorbing and gain media.

\section{\label{sec:level1}References\protect\\  \lowercase{} }

[1] John David Jackson, Classical Electrodynamics, 1975 pp.313-326. 

[2] Abramowitz, M. and Stegun, I. A. (Eds.). "Circular Functions." §4.3 in Handbook of Mathematical Functions with Formulas, Graphs, and Mathematical Tables, 9th printing. New York: Dover, p. 75, 1972. 

[3] Brillouin, Wave Propagation and Group Velocity, 1960. pp.23-35.

[4] Gain assisted superluminal light propagation,Wang etal. Nature 406, 277-279 (20 July 2000).

[5] Linear Pulse Propagation in an absorbing medium, Chu etal. Phys. Rev. Lett. 48, 738 – Published 15 March 1982.

[6]  http://arxiv.org/abs/1001.0249

\end{document}